\documentclass[aps,prl,twocolumn,showpacs,floatfix,a4paper]{revtex4}
\usepackage{graphicx,amsfonts}
\usepackage[intlimits]{amsmath}

\begin{document}

\title{Entanglement entropy of low-lying excitation in localized interacting 
system: Signature of Fock space delocalization}
\author{Richard Berkovits}
\affiliation{Department of Physics, Bar-Ilan
University, Ramat-Gan 52900, Israel}

\begin{abstract}
The properties of the entanglement entropy (EE) of low-lying excitations
in one-dimensional disordered interacting systems are studied. 
The ground state EE shows a clear signature of localization, while
low-lying excitation shows a crossover from metallic behavior at short
sample sizes to localized at longer length. The dependence of the
crossover as function of interaction strength and sample length is
studied using the density matrix renormalization group (DMRG). 
This behavior corresponds to the presence of the predicted many particle
critical energy in the vicinity of the Fermi energy. 
Implications of these results to experiments are
discussed.
\end{abstract}


\maketitle

Many body localization (MBL) has drawn growing interest
in recent years
\cite{AGKL97,gornyi05,basko06,berkovits98,oganesyan07,znidaric08,aizenman09,monthus10,berkelbach10,pal10,canovi11,cuevas12,bardarson12,serbyn13,vosk13,iyer13,huse13,huse13a,serbyn13a,bauer13}.
The Anderson localization transition \cite{anderson58,lee85}
is a zero-temperature quantum phase 
transition, which for the non-interacting case is manifested  
in the properties of the single-particle eigenstates
and eigenvalues \cite{shklovskii93}. For one and two dimensional
systems all the single-particle state are localized for 
any amount of disorder, while for 
three dimensional systems at a given disorder there exists a critical 
energy (mobility edge) below which all states are localized.

For a many-particle system, as long as no particle-particle interactions
are present, the properties of a many-particle excited state are determined
by the properties of the single-particle states. Thus, as long as all 
the particles occupy localized single-particle states 
the many-particle states should
exhibit localized behavior. For example, the entanglement entropy (EE)
between any 
sub-region A of the system and the rest of it should saturate
(i.e., $S_A \propto \xi^{d-1}$, where $d$ is the dimensionality of the system
and $\xi$ is the single-electron localization length).
While if some of the extended single-particle states are occupied
the usual volume law for the EE ($S_A \propto L_A^d$, where $L_A$ is the 
length of region A) is observed 
\cite{huse13}. Physically, the difference between the localized
and extended many-particle states is manifested in the dynamics of
a particle tunneling into some particular region of the system, or a particle
excited by a confined external perturbation in the region. In the localized
phase it will remain there 
(i.e., will have a very long lifetime -- sharp level),
while in the extended case it will leave the region (short lifetime -- 
broad level). 

What is the influence of particle-particle 
interaction on the above picture?
A cursory consideration may lead to the conclusion that
since particles now interact with each other, any localized excitation
will eventually spread all over the system. Thus,
the rest of the system acts as a thermal bath for any excited
sub-system \cite{deutsch91, srednicki94, rigol08}.
Surprisingly, Basko, Aleiner and Altshuler \cite{basko06}
have shown that if all the single-electron states are localized,
an excitation will remain localized even in the presence of 
particle-particle interactions up to a critical temperature
or excitation energy. This behavior is known as the MBL.

One can expect to see a signature of this behavior also in the
EE between a sub-region and the rest of the system. For excitations
below the the MBL mobility edge the EE should follow its localized
behavior. This was numerically demonstrated for the
ground state of interacting electrons in a 1D disordered system 
\cite{berkovits12}, where repulsive interactions only bolsters the localization
behavior. On the other hand, for excited states above the mobility edge 
the EE should follow the volume law expected from thermalized states.

Let us examine the parametric dependence of $T_C$ on the interaction
strength more carefully.
The critical
temperature $T_C \sim \Delta_{\xi}/ \lambda|\ln \lambda|$ 
\cite{basko06}
where $\Delta_{\xi}$ is the
single level spacing in a region of size $\xi$,
and $\lambda$ is the dimensionless interaction strength 
proportional
to $U_{\rm max}/\Delta_{\xi}$, where $U_{\rm max}$ is the 
maximal matrix element
between two many-particle states proportional to $U(a/\xi)^d$. Here
$U$ is the strength of the nearest-neighbor interaction, 
$a$ is the range of interaction and $d$ the 
dimensionality. The level spacing
in the localization volume $\Delta_{\xi}=1/(\nu \xi^d)$ , where $\nu$ is the 
density of states. 
When the critical temperature is translated to a critical excitation energy
high in the excitation spectra, 
${\mathcal E}_C= n d T_C$, where $n$ is the number of particles excited.
At a constant filling $n \propto L^d$, where $L$ is the system size, and
therefore ${\mathcal E}_C$ is extensive. Thus, for a given system size,
${\mathcal E}_C$ decreases as $U$ increases,
while ${\mathcal E}_C$ increases as $L$ increases.

Probing the transition at low excitation energies is advantageous both for 
experimental and numerical reasons. Experimentally, at lower excitation
energies the influence of electron-phonon scattering (which will thermalize
even in the MBL regime) is weaker, while
numerically it is possible to used the accurate DMRG method only close to
the ground state. Of course, strictly speaking finite systems do not
exhibit a true phase transition, but they can nevertheless show a signature
of such a transition and reveal its properties.

For a finite system of length
$L$ the spacing between the ground state and the first many-particle excited 
state is $\Delta_1 = 1/ \nu L^d$. At low energies
only a few (order of one) quasi-particles are excited and 
${\mathcal E}_C \sim d T_C$, thus intensive. 
One expects a crossover in the behavior of the low-lying excitations
from a metallic behavior once $\Delta_1 > {\mathcal E}_C$ 
to localized when $\Delta_1 < {\mathcal E}_C$. 
Thus, for a finite disordered system 
all low lying excited states will exhibit localization 
as long as $L>(\lambda |\ln(\lambda)|)^{1/d} \xi$. 
Of course, localization will only occur on length scales
larger than $L>\xi$.
Since in all these considerations one assumes $\lambda \le 1$  
one would expect that the length scale for which the excited state
will exhibit localization behavior ($L>\xi$) 
is of order of the single-electron localization length.

The above consideration leaves out an important fact
regarding the ground-state, namely that the ground-state many-particle
localization length of a typical 
disordered system is strongly suppressed by repulsive interactions.
For example, a spinless 1D wire of length $L$,
with on-site disorder and nearest neighbor interactions described by
the following Hamiltonian:
\begin{eqnarray} \label{hamiltonian}
H &=& 
\displaystyle \sum_{j=1}^{L} \epsilon_j {\hat c}^{\dagger}_{j}{\hat c}_{j}
-t \displaystyle \sum_{j=1}^{L-1}({\hat c}^{\dagger}_{j}{\hat c}_{j+1} + h.c.) 
\\ \nonumber
&+& U \displaystyle \sum_{j=1}^{L-1}({\hat c}^{\dagger}_{j}{\hat c}_{j} 
- \frac{1}{2})
({\hat c}^{\dagger}_{j+1}{\hat c}_{j+1} - \frac{1}{2}),
\end{eqnarray}
where $\epsilon_j$ is the random on-site energy, taken from a uniform 
distribution in the range $[-W/2,W/2]$,
$U$ is the interaction strength, and $t=1$ is the
hopping matrix element.
${\hat c}_j^{\dagger}$ is the creation 
operator of a spinless electron at site $j$ in the wire, and
a neutralizing background is included in the interaction term.
This model 
is localized for non-interacting as well as any repulsive $U>0$ 
interactions \cite{apel_82,giamarchi88}, and has a metallic regime for
attractive interaction in the vicinity of $U=-1$ and not to
strong disorder
\cite{schmitt98,schuster02,carter05,chu13}.
For the non-interacting case close to half-filling the 
single-particle localization
length, $\xi(W,U=0) \approx 105/W^2$
\cite{romer97}. 
Once particle-particle interactions are included, properties
of the system such as transport (and entanglement) are
determined by the many-particle wave function and the
many-particle localization length deviates
from the single-particle localization length.
The many-particle localization length is related to the
single-particle localization length by \cite{apel_82,giamarchi88}
$\xi_{MP}=\xi(W,U) = (\xi(W,U=0))^{1/(3-2g(U))}$,
where $g(U)=\pi/[2 \cos ^{-1} (-U/2)]$ 
is the Luttinger parameter \cite{g_formula}. 
Thus, for the non-interacting case $g=1$ and $\xi_{MP}=\xi$ while
for $U=-1$, $g=3/2$ and $\xi_{MP}$ diverges. 

For repulsive interactions $\xi_{MP}<\xi$ since $g(U>0) < 1$,
and the length scales for which the ground state of a finite system
will exhibit metallic behavior becomes shorter ($L<\xi_{MP}<\xi$).
This has been verified  by analyzing the behavior of the EE
of the ground state \cite{berkovits12}. 
On the other hand, for the low-lying excitations
according to Basko, Aleiner, and Altshuler's argument \cite{basko06}
the excitations are expected to remain metallic for
$L<O (\xi)$.
Thus, for interacting systems a significant range of $L$ 
(roughly estimated as $\xi_{MP}<L<\xi$) for which
{\it all} low lying excited states will exhibit metallic
behavior while the ground state is localized may be expected.

In this letter we study the behavior of low-lying excitations of a one-dimensional 
finite interacting disordered system 
and demonstrate the existence of a size-regime for which the system ground state 
is localized while any excited state is metallic.
Unlike when studying high excitations for which there are no efficient
numerical methods and 
one is reduced to treating small systems, simplify the system or making
other assumptions regarding the solutions \cite{berkovits98,oganesyan07,monthus10,berkelbach10,pal10,canovi11,cuevas12,bardarson12}, 
the density matrix 
renormalization group (DMRG) numerical method is very accurate for  
low-lying excitations and can handle large systems.

In order to identify whether an excited state is metallic or localized we shall
use its EE, which for the ground state has been shown to be a very accurate
way to determine the localization length \cite{berkovits12}
and metal-insulator transitions point \cite{chu13}, for interacting
disordered system. Nonetheless, unlike the ground state EE (GSEE) which is well
understood, the EE behavior of low lying-excitations needs further
clarification \cite{alcaraz08,masanes09,berganza12,taddia13}. 
We shall begin by studying the EE of low lying-excitations in
the metallic regime before proceeding to the localized regime.

\begin{figure}
\includegraphics[width=8.3cm,height=!]{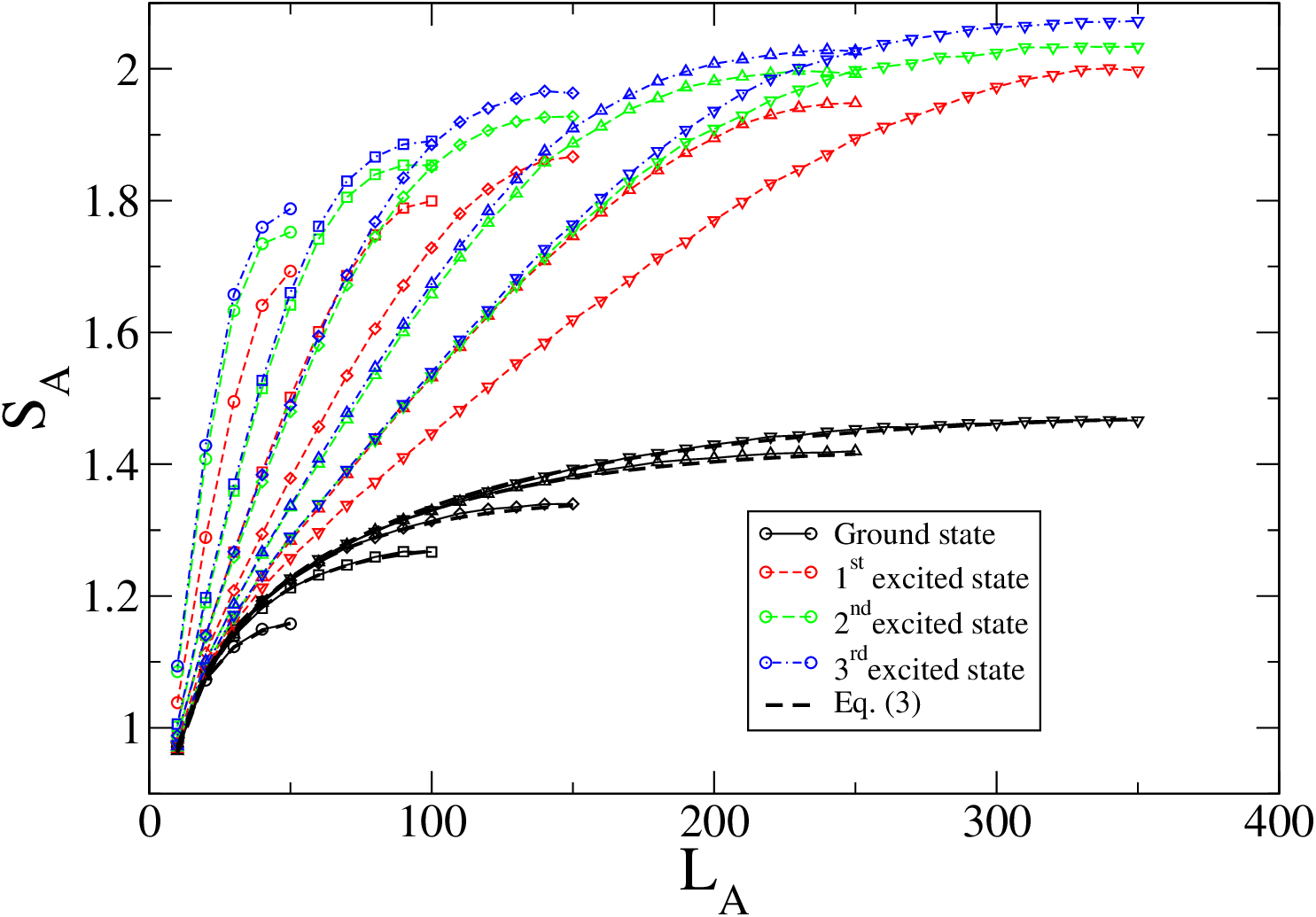}
\vskip 1.1truecm
\includegraphics[width=8.3cm,height=!]{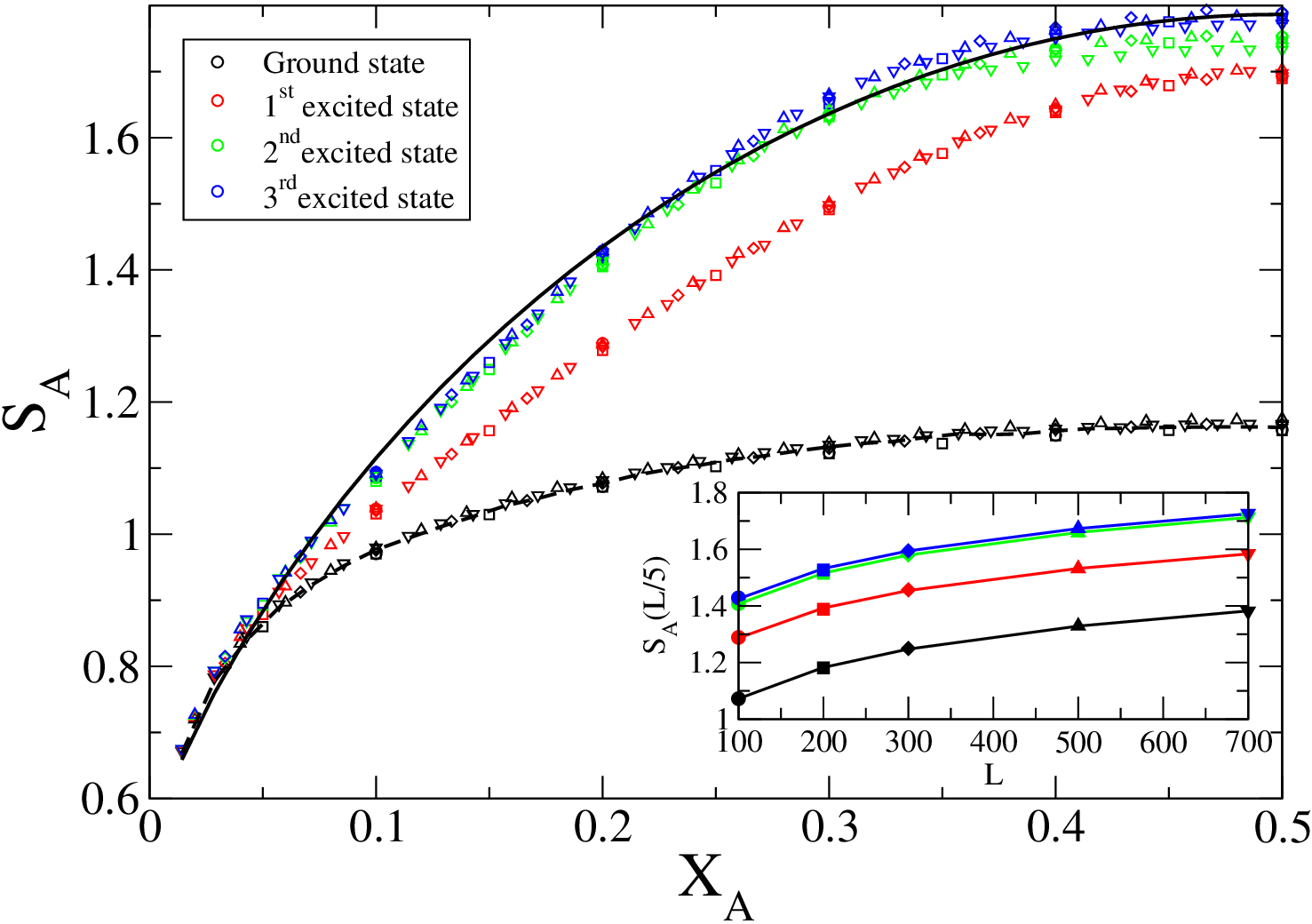}
\caption{\label{fig1}
(Color online)
Upper panel:
The EE of a metallic system as function of $L_A$,
for different system sizes $L$.
The symbols correspond to the DMRG results, where the different sample length
are represented by different symbols $\bigcirc,L=100$; $\square,L=200$;
$\Diamond,L=300$; $\bigtriangleup,L=500$; and $\bigtriangledown,L=700$.
The colors correspond to the excitation number (continuous black 
- ground state, dashed red - first excitation,
long dashed green - second excitation, dot-dashed blue - third excitation).
The thick dashed line correspond to Eq. (\ref{ee_erf}).
Lower panel:
The EE scaled as function of the ratio $X_A=L_A/L$.
The thick line correspond to Eq. (\ref{esee}) with $n_0=1.82$.
Inset: The EE of the ground state and the low lying excitations 
for $L_A=0.2 L$ as function of $L$. The  symbols correspond to the
numerically calculated EE, while the lines to a fit to  $c^{(i)}+(1/6)\ln(L)$,
where $c^{(i)}$ depends only on the excitation number.
}
\end{figure}

The EE of a pure state $|\Psi\rangle$ in a sample partitioned into
two sections A and B of length $L_A$ and $L_B=L-L_A$ is given
by
\begin{eqnarray}
S_{A/B}=-{\rm Tr} \rho_{A/B} \ln \rho_{A/B},
\label{sa}
\end{eqnarray}
where the reduced density matrix 
$\rho_{A/B}={\rm Tr}_{B/A}|\Psi\rangle\langle \Psi |$.
The ground-state and the three lowest excited states are calculated.
As usual in DMRG, the reduced density matrix for each state at each length
$L_A$ is calculated and diagonalized, thus obtaining the EE does not involve
any additional numerical overhead. 
First we present the EE in the metallic regime, i.e.,
$\xi_{MB}\rightarrow \infty$ for $W=0.7$ ($\xi \sim 210$), and $U=-1$.
The median EE $S(L_A,L)$ for all four states 
over $100$ realizations of disorder are
plotted as function of $L_A$ in 
the upper panel of Fig. \ref{fig1}, for
system length $L=100$,$200$,$300$,$500$ and $700$.

It is clear that the ground state in the metallic regime follows the 
expected logarithmic behavior \cite{holzhey94,vidal03,calabrese04,korepin04}
\begin{eqnarray} \label{ee_erf}
S_{\rm gs}(L_A,L) = \frac{1}{6}
\ln\left(\sin\left(\pi X_A\right)
\right)+c_g,
\end{eqnarray} 
where $X_A=L_A/L$ and $c_g$ is a non-universal 
constant. 
Using Eq. (\ref{ee_erf}) results in a perfect fit for GSEE.
The excited state EE (ESEE) has a different functional form. Nevertheless,
the only relevant scale remains $X_A=L_A/L$, and therefore we expect
the ESEE to follow: 
\begin{eqnarray} \label{ee_ex}
S^{(i)}(L_A,L) = 
S^{(i)}(X_A) + c^{(i)} (L).
\end{eqnarray}
The superscript $(i)$ denotes the excitation number.
As can be seen in the lower panel of Fig. \ref{fig1}, this scaling leads
to a collapse of $S^{(i)}(X_A)$ on the same curve
for different values of $L$.
The constant $c^{(i)}(L)$ is chosen so 
$S^{(i)}(X_A=0.2)$ is equal for different $L$. As can be seen from the
inset in Fig. \ref{fig1}, where $S^{(i)}(L_A=0.2L,L)$
for different values of $L$ is plotted,  $c^{(i)}(L) = c^{(i)}+(1/6)\ln(L)$,
where $c^{(i)}$ depends only on the excitation number.

Unlike the GSEE, for the ESEE there is no
established functional form for the behavior of 
$S^{(i)}(X_A)$. Nevertheless, using
the form of the typical EE of several excited particles 
in a clean system \cite{berkovits13} as a guide, 
one may try to fit the ESEE to
\begin{eqnarray} \label{esee}
S^{(i)}(X_A)=-n_0(X_A \ln X_A + (1-X_A) \ln(1-X_A))+c^{(i)},
\end{eqnarray} 
where,
$n_0$, for a clean system is the number of excited particles, while here
it is treated as a fitting parameter. One might understand this functional
form by considering that the ESEE should follow 
the volume law (up to logarithmic corrections
appearing also in the ground state). 
For the second and third excitation
this crude analogy gives a reasonable description with $n_0=1.82$ of the EE. 

\begin{figure}
\includegraphics[width=8.3cm,height=!]{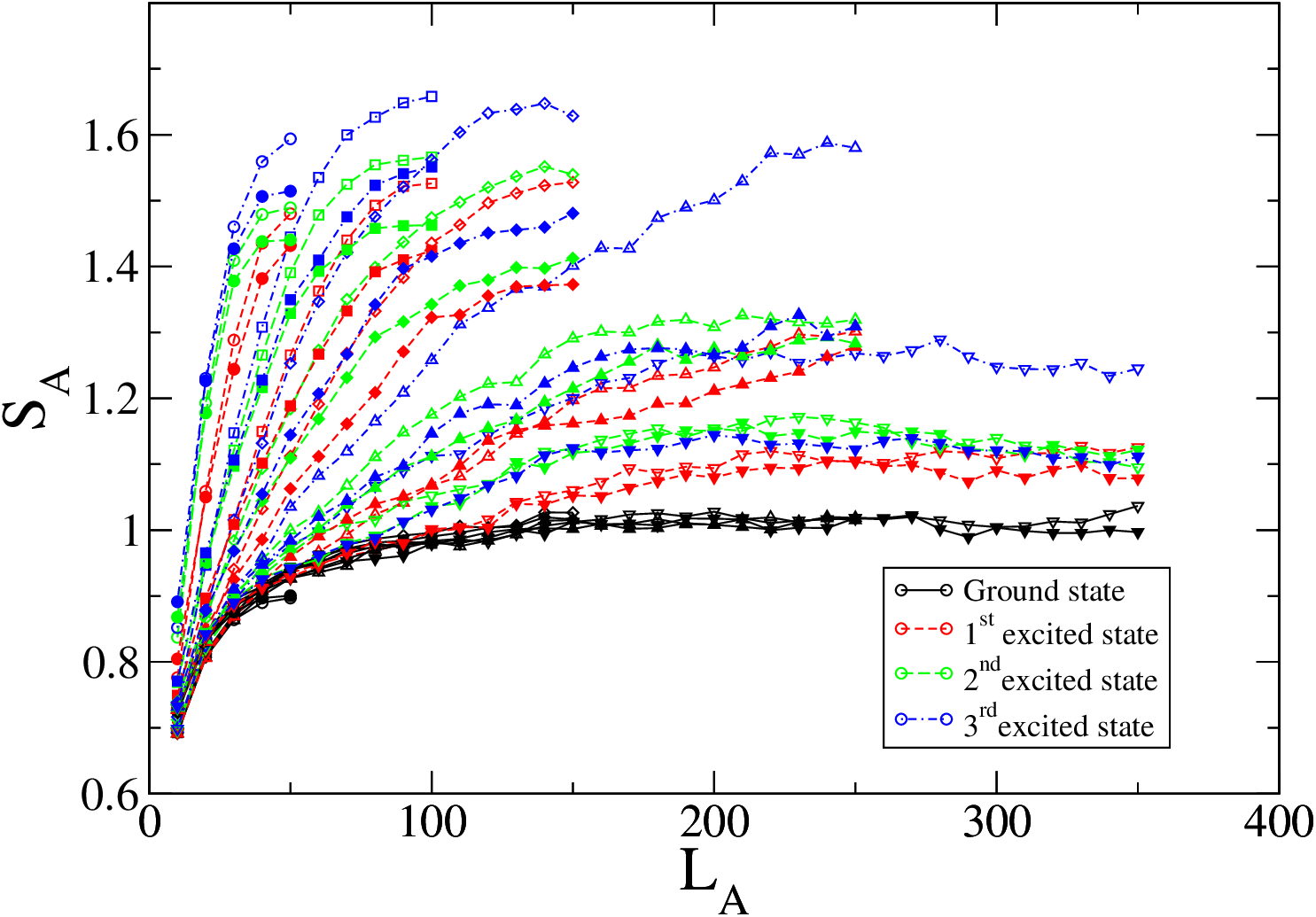}
\vskip 1.1truecm
\includegraphics[width=8.3cm,height=!]{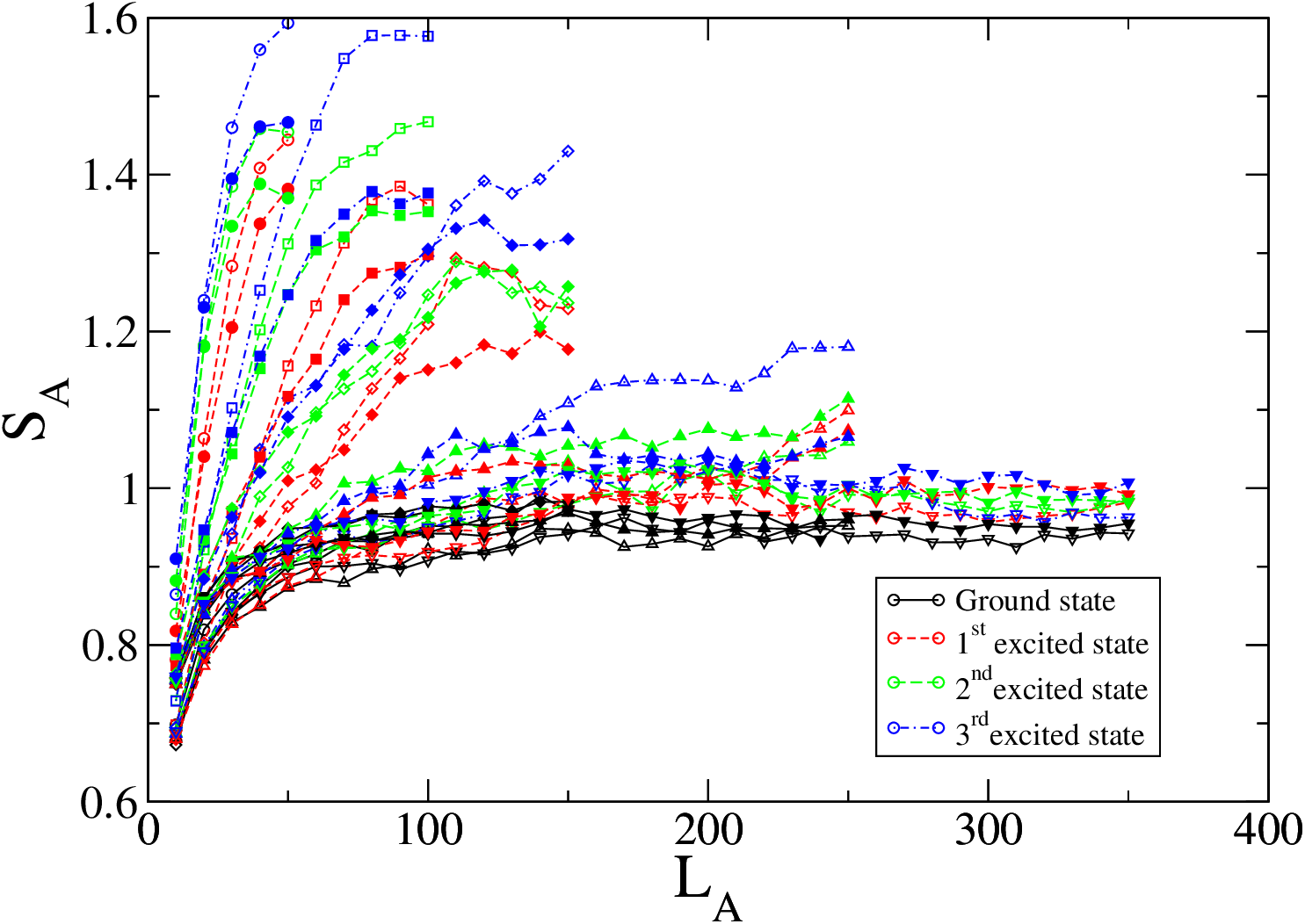}
\caption{\label{fig2}
(Color online)
The EE of a weakly disordered system (upper panel, $\xi_{MP} \sim 100$)
and strongly disordered systems (lower panel, $\xi_{MP} \sim 50$) 
as function of $L_A$, for different system sizes $L$.
The symbols represent DMRG results for different 
sample length as in Fig. \ref{fig1}.
The different excitations are indicated by color and line-shape.
Upper panel:
The full symbols correspond to $U=0.6$ and $W=0.5$ 
($\xi\sim 400,\xi_{MP} \sim 100$), while
empty symbols to $U=1.8$ and $W=0.15$
($\xi\sim 4500,\xi_{MP} \sim 100$).
Lower panel:
The full symbols correspond to $U=0.6$ and $W=0.7$ 
($\xi\sim 200,\xi_{MP} \sim 50$), while
empty symbols to $U=1.8$ and $W=0.25$
($\xi\sim 1600,\xi_{MP} \sim 50$).
}
\end{figure}

How does disorder change the low-lying ESEE? 
For the ground state the logarithmic correction saturates on
the scale of $\xi_{MP}$, resulting in a clear signature of
many particle localization\cite{berkovits12}.
Indeed, this is seen in the behavior of the GSEE
depicted in Fig. \ref{fig2}, where the EE for systems with different
single-electron localization length, but equal many-particle 
localization length fall on top each other for $L>\xi_{MP}$.  
On the other hand, the ESEE shows a different behavior.
For short length scales (albeit longer than $\xi_{MP}$) the
low-lying excitations behave differently than the ground state 
and continue to follow the volume law typical for the ESEE in 
the metallic regime (Fig. \ref{fig1}). Only for larger system sizes
does the low-lying ESEE begin to resemble the GSEE.

We expect the volume law to
be strongly affected by localization. For system sizes much shorter than
the localization length the EE should not be seriously affected,
while for sizes larger than the localization length it should saturate
and resemble the GSEE.
Indeed, this crossover in the behavior of the ESEE
is clearly seen in Fig. \ref{fig2}. For the weak disorder case 
($\xi_{MP} \sim 100$), the
EE shows the volume law to hold up until $L=300$ (even up to $L=500$, for the
third excited state of strong interactions $U=1.8$), 
while crossing over to the GSEE
behavior at $L=700$. For stronger disorder ($\xi_{MP} \sim 50$), 
the volume law behavior is seen up
to $L=200$, while at $L=500$ the ESEE is already closer
to an area law. Thus, the GSEE shows localized behavior corresponding
to $\xi_{MP}$, while the low-lying ESEE show volume law (metallic)
behavior on larger length scales. Moreover, the crossover to localized
behavior occurs at larger sample sizes for higher $U$ and higher excitations,
as expected from the many-body localization scenario.

\begin{figure}
\includegraphics[width=8.5cm,height=!]{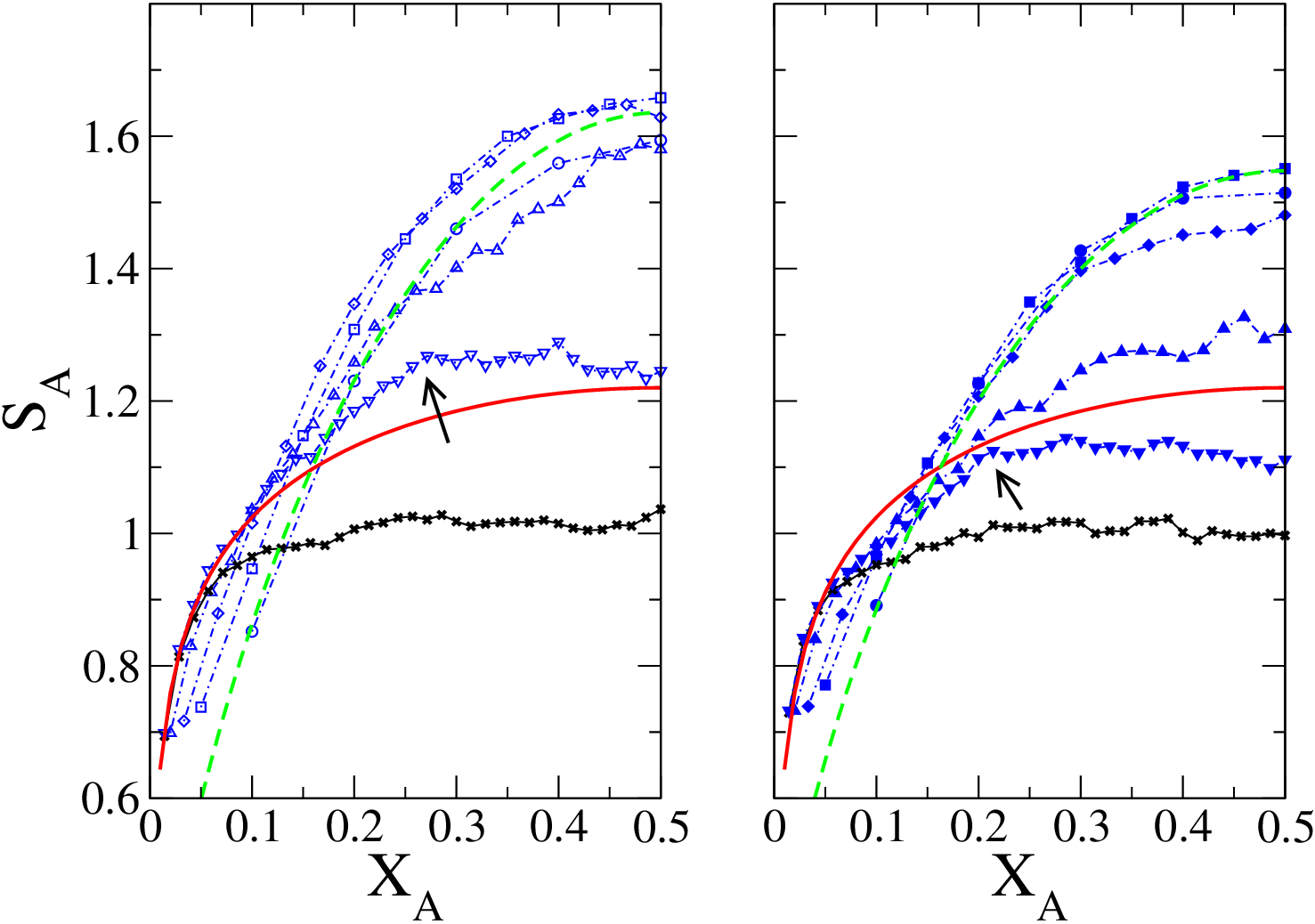}
\caption{\label{fig3}
(Color online)
The third ESEE for the weakly disordered system 
($\xi_{MP} \sim 100$) depicted in the upper panel of Fig. \ref{fig2}
as function of $X_A$. Again, the symbols correspond to
the DMRG results as in Fig. \ref{fig1}.
For comparison the EE of the ground state of $L=700$ is depicted
by black X symbols. The arrows indicate the point of saturation of the
EE for $L=700$.
(Left panel)
Strong interactions ($U=1.8$).
The dashed green curve corresponds to Eq. (\ref{esee}) with $n_0=2.1$ and $c^{(i)}=0.18$,
while the red curve corresponds to Eq. (\ref{ee_erf}) with $c_g=1.22$.
(Right panel)
Weak interactions ($U=0.6$)
The dashed green curve corresponds to  Eq. (\ref{esee}) with $n_0=1.8$ and $c^{(i)}=0.3$, 
and the red curve corresponds to Eq. (\ref{ee_erf}) with $c_g=1.22$.
}
\end{figure}

Let us examine the behavior of the ESEE more carefully.
As we have seen,
for the clean case the ESEE scales as $X_A=L_A/L$.
and follows a volume law behavior (Eq. (\ref{esee})), while
the ground state follows the area law (Eq. (\ref{ee_erf})). 
For $L\ll\xi^{(i)}$
(where $\xi^{(i)}$ is the localization length of the $i$-th excited state)
the ESEE is expected to follow the volume law. Deep in the localized regime
($L \gg \xi^{(i)}$) there should be no difference between the ESEE 
and the GSEE.
Thus, contrary to the situation for the GSEE where $\xi$
plays the role of a saturation length in the area law, for the 
ESEE it also changes the functional form of the EE.
This is clearly seen in Fig. \ref{fig3}, where the behavior
of the 3rd excited state for the weak disorder is depicted. For
$L=100$ the behavior for both interaction strength fit a volume law.
At larger length a crossover toward an area law behavior is seen.
This crossover has two distinct features. There is saturation of the EE
at large $X_A$, which is clear for $L=700$ for both $U=1.8$ and $U=0.6$. At weak
interaction $U=0.6$ this saturation is clear also for shorter sample sizes 
($L=500$). This saturation is similar to the ground state
EE saturation, but for the ESEE it is accompanied by
a change in the behavior at small values of $X_A$. 
Again, for $L=700$ for both $U=1.8$ and $U=0.6$ the ESEE
is indistinguishable for small values of $X_A$ from the GSEE.
Thus, unlike for the GSEE, the
effect of the localization crossover is seen on length scale smaller
than $\xi$, and involve a change in the functional behavior of the ESEE. 

It is obvious from Fig. \ref{fig3} that the description of the EE in
the crossover regime is not trivial, and there is no straight forward 
way to extract
the excited state localization length, $\xi^{(i)}$. 
When $\xi^{(3)}$ is larger than $L$, The ESEE follows metallic
behavior (Eq. (\ref{esee})), and therefore its clear that $\xi^{(3)}>500$
as long as $L<500$ for $U=1.8$ and $\xi^{(3)}>300$
as long as $L<300$ for $U=0.6$ (compared to $\xi_{MP}=100$).
As the samples are longer, $\xi^{(3)}$ becomes shorter. At   
sample length for which the ESEE behaves
similar to the GSEE for small values of $X_A$, one may roughly extract the
the localization length by identifying the saturation of the ESEE. 
For sample length of $L=700$,  $\xi^{(3)}$ is indicated
by an arrow and corresponds to $\xi^{(3)}\sim 175$ for $U=1.8$ and 
$\xi^{(3)}\sim 140$ for $U=0.6$. 
Thus even at sample length of $L=700$, the the 3rd excited state localization
is significantly larger than the ground state localization length
$\xi_{MP}=100$.

How could this enhancement of the localization length of the low-lying excited
states be experimentally verified for small finite samples? 
In order to distinguish between the ground state and the low lying excitations
one should have systems small enough so the energy gap between
the ground state and low lying excited states is larger than
other experimental energy scales such as temperature or source-drain voltage
($\Delta_1 > T,V_{SD}$), but larger than the many-particle localization length. 
By using photons to excite the low-lying excited
states one could hope to observe a strong (even orders of magnitude) enhancement
of the photoconductivity for these small samples, or to probe
the conductance optically \cite{zvi}. In order to rule
out other effects which may enhance the photoconductivity such as heating, 
the effect should strongly decrease for larger samples.

In conclusion, For short samples, low lying excited states may show metallic
behavior although the sample is much longer than the ground state localization
length and the ground state is strongly localized. Only for longer systems
crossover to the localized regime occurs. This is a clear signature of the MBL
mobility edge which can approach the Fermi energy for short systems. The
behavior of the low lying excitation for short systems may be probed using
optical techniques.


\begin{acknowledgments}
I would like to thank Z. Ovadyahu and I. L. Aleiner for useful discussions.
Financial support from the Israel Science Foundation (Grant 686/10) is
gratefully acknowledged.
\end{acknowledgments}

\end{document}